\documentclass[12pt]{article}

\usepackage[totalheight = 23cm, totalwidth = 17cm]{geometry}
\usepackage{amssymb,amsmath,amsfonts,amsbsy,epsfig}

\begin{document}

\begin{titlepage}

\begin{center}

\vspace*{-10ex}
\hspace*{\fill}
YITP-07-68

\vskip 1.5cm

\Huge{The lessons from the running of the tensor-to-scalar ratio}

\vskip 1cm

\large{Jinn-Ouk Gong$^{1,2,3}$\footnote{jgong\_AT\_ hep.wisc.edu}
\\
\vspace{0.5cm}
{\em
 ${}^1$ Harish-Chandra Research Institute
 \\
 Chhatnag Road, Jhunsi, Allahabad, 211 019, India
 \\
 \vspace{0.2cm}
 ${}^2$ Yukawa Institute for Theoretical Physics
 \\
 Kyoto University, Kyoto 606-8502, Japan
 \\
 \vspace{0.2cm}
 ${}^3$ Department of Physics, University of Wisconsin-Madison
 \\1150 University Avenue, Madison, WI 53706-1390, USA
}}

\vskip 0.5cm

\today

\vskip 1.2cm

\end{center}

\begin{abstract}

We derive a simple consistency relation from the running of the tensor-to-scalar
ratio. This new relation is first order in the slow-roll approximation. While for
single field models we can obtain what can be found by using other observables,
multi-field cases in general give non-trivial contributions dependent on the
geometry of the field space and the inflationary dynamics, which can be probed
observationally from this relation. The running of the tensor-to-scalar ratio may be
detected by direct laser interferometer experiments.

\end{abstract}

\end{titlepage}

\setcounter{page}{0}
\newpage
\setcounter{page}{1}

Inflation~\cite{inflation} is supposed to be the most promising candidate to solve
many cosmological problems such as horizon problem, and to provide the initial
conditions for the subsequent standard hot big bang universe~\cite{books}. Among
many successes of inflation, the nearly scale invariant density perturbations on the
scales of the cosmic microwave background (CMB) are believed to be one of the most
solid predictions of inflation. They were first probed by the Cosmic Background
Explorer (COBE) satellite~\cite{cobe}, and indeed in excellent agreement with most
recent observations~\cite{observations}. In the standard picture, the origin of
these scale invariant perturbations is the vacuum fluctuations of one or more scalar
fields which dominate the energy density of the universe during inflation, the
inflaton fields.

There exists, however, another source of the CMB temperature anisotropies produced
during inflation. The tensor perturbations, which emerge as the primordial
gravitational waves~\cite{gw}, are believed to contribute to the CMB anisotropies at
low multipoles up to $l \lesssim 100$. There are several reasons why the
gravitational waves, and the detection of them, are potentially very important. One
reason should be that from the amplitude of the gravitational waves, or the
tensor-to-scalar ratio $r \equiv \mathcal{P}_\mathrm{T}/\mathcal{P}_\mathrm{S}$, we
can directly determine the inflationary energy scale~\cite{infE}. Another reason
which becomes popular recently is that, in many models of string inflation based on
the de Sitter vacua constructed by flux compactification~\cite{kklt}, the level of
the gravitational waves are expected to be extremely small to be ever detected, $r
\lesssim 10^{-24}$ or so~\cite{stringgw}. Meanwhile the current bound is only $r
\lesssim 0.2$. In single field models, detectable gravitational waves means a field
variation of $\mathcal{O}(m_\mathrm{Pl})$ with $m_\mathrm{Pl} \equiv (8\pi G)^{-1/2}
\approx 2.4 \times 10^{18}\mathrm{GeV}$~\cite{lythbound}. Constructing a stringy
model of inflation with such a large field variation and in turn detectable
gravitational waves remains an open challenge \cite{largeGW}. Moreover, it is
expected that by near future CMB observations the sensitivity will be improved at
most to a level of $r \sim \mathcal{O}\left( 10^{-2} \right)$. Thus it is very
important to fully understand the implications of the primordial gravitational waves
regarding the observations.

In this note we explore another aspect of the tensor perturbations. We point out
that we can obtain a new consistency relation from the running of the
tensor-to-scalar ratio. Using this relation, we might be enabled to observationally
discriminate multi-field models from single field ones. More importantly, for
multi-field cases, this consistency relation opens a new possibility to study the
geometry of the field space and the dynamics during inflation.

Irrespective of inflation model driven by single or multiple number of fields, the
spectrum of the primordial gravitational waves produced during inflation is given
by~\cite{gw}
\begin{equation}\label{gwP}
\mathcal{P}_\mathrm{T} = \frac{8}{m_\mathrm{Pl}^2} \left( \frac{H}{2\pi} \right)^2
\, ,
\end{equation}
and the dependence on the wavenumber $k$ is described by
\begin{equation}\label{gwPkdep}
\mathcal{P}_\mathrm{T} \propto k^{n_\mathrm{T}} \, ,
\end{equation}
where the corresponding spectral index $n_\mathrm{T}$ is
\begin{equation}\label{gwindex}
n_\mathrm{T} = -2\epsilon \, ,
\end{equation}
with $\epsilon \equiv -\dot{H}/H^2$ being the usual slow-roll parameter. This is
because the equation of motion of each polarization state of the gravitational waves
is exactly that of massless scalar field~\cite{gwcalc} and is thus independent of
the detail of the inflation model under consideration.

Meanwhile, the spectrum and the spectral index of scalar perturbations do depend on
the detailed inflationary dynamics\footnote{Note that if we allow post-inflationary
generation of perturbation via e.g. the curvaton mechanism~\cite{curvaton},
$\mathcal{P}_\mathrm{S}$ occupies only some or even negligible fraction of the total
scalar perturbations, and in turn we always obtain smaller tensor-to-scalar
ratio~\cite{postinfr}. This jeopardizes the possibility of using the `standard'
consistency relation $r = 16\epsilon$ to distinguish between single and multi-field
inflation models. In the present note, however, we do not consider any
post-inflationary production of perturbation.}. More specifically, we have
\begin{equation}\label{singleP}
\mathcal{P}_\mathrm{S} = \left( \frac{H}{2\pi} \right)^2 \left( \frac{H}{\dot\phi}
\right)^2 \, ,
\end{equation}
and
\begin{equation}\label{multiP}
\mathcal{P}_\mathrm{S} = \left( \frac{H}{2\pi} \right)^2 h^{ij} N_{,i}N_{,j} \, ,
\end{equation}
for the cases of single and multiple fields, respectively. In Eq.~(\ref{multiP}),
$h_{ij}$ is the metric of the field space, $N \equiv \int H dt$ is the number of
$e$-folds, and $N_{,i} \equiv \partial{N}/\partial\phi^i$. The corresponding
spectral indices which will be shown soon are different, but nevertheless we can
write the dependence of $\mathcal{P}_\mathrm{S}$ on $k$ as
\begin{equation}\label{scalarPindex}
\mathcal{P}_\mathrm{S} \propto k^{n_\mathrm{S}-1} \, .
\end{equation}
Combining Eqs.~(\ref{gwPkdep}) and (\ref{scalarPindex}), $r$ depends on $k$ as
\begin{equation}
r \propto k^{1 - n_\mathrm{S} + n_\mathrm{T}} \, ,
\end{equation}
so that the running of $r$ is
\begin{equation}\label{rrunning}
\frac{d\log r}{d\log k} = 1 - n_\mathrm{S} + n_\mathrm{T} \, .
\end{equation}
We stress that this relation is valid for all the inflationary models, irrespective
of driven by single or multiple number of fields. Moreover, it is first order in the
slow-roll approximation and does not involve any model dependent parameter. The
implications Eq.~(\ref{rrunning}) suggests for single and multi-field inflation,
however, are quite different.

For single field case, it is well known that the spectral index is\footnote{It is
noticeable that while one can proceed the calculation with corrections up to
arbitrary power of the slow-roll parameters~\cite{singlecorr}, more general
slow-roll conditions~\cite{generalSR} may give rise to, for example, large enough
running of the scalar spectral index. This is another observationally interesting
possibility.}
\begin{equation}\label{singlePindex}
n_\mathrm{S} - 1 = -6\epsilon + 2\eta \, ,
\end{equation}
where
\begin{equation}\label{singleeta}
\eta \equiv m_\mathrm{Pl}^2 \frac{V''}{V} \, ,
\end{equation}
is another slow-roll parameter. Thus, from Eq.~(\ref{rrunning}),
\begin{equation}\label{singlerrunning}
\frac{d\log r}{d\log k} = 2 (2\epsilon - \eta) \, .
\end{equation}
Note that the same factor of $2\epsilon - \eta$ can be derived from a higher order
version of the consistency relation~\cite{hierarchy}. The relation here is, however,
first order in the slow-roll parameters as we emphasized. This also provides some
information in the light of a classification scheme of the models of
inflation~\cite{zoology}. However, we can extract exactly the same conclusion by
combining $\mathcal{P}_\mathrm{S}$, $\mathcal{P}_\mathrm{T}$ (or $r$) and
$n_\mathrm{S}$. Measuring $d\log r/d\log k$ will give a consistency check on the
single field model and therefore may be worth in that sense, but there is nothing
new.

The situation is, however, completely different for multi-field inflation models.
When multiple number of light fields give rise to inflation, following $\delta{N}$
formalism~\cite{deltaN}, the spectral index is given by
\begin{equation}\label{multiPindex}
n_\mathrm{S} - 1 = -2\epsilon - \frac{r}{4} + 2\eta_\mathrm{multi} -
2\frac{N_{,i}N_{,j}}{h^{kl}N_{,k}N_{,l}}
\frac{{R^i_{\,\,ab}}^j}{3m_\mathrm{Pl}^2}\frac{\dot\phi^a\dot\phi^b}{H^2} \, ,
\end{equation}
where $R^i_{\,jkl}$ is the Riemann curvature tensor of the field space,
\begin{equation}\label{riemann}
R^i_{\,jkl} = \Gamma^i_{jk,l} - \Gamma^i_{jl,k} + \Gamma^m_{jk}\Gamma^i_{lm} -
\Gamma^m_{jl}\Gamma^i_{km} \, ,
\end{equation}
with
\begin{equation}\label{christoffel}
\Gamma^i_{jk} \equiv \frac{1}{2}h^{il} \left( h_{jl,k} + h_{kl,j} - h_{jk,l} \right)
\end{equation}
being the Christoffel symbol constructed by $h_{ij}$, and we have defined, {\`a} la
Eq.~(\ref{singleeta}),
\begin{equation}
\eta_\mathrm{multi} \equiv m_\mathrm{Pl}^2 \frac{N_{,i}N_{,j}}{h^{kl}N_{,k}N_{,l}}
\frac{V^{;ij}}{V} \, ,
\end{equation}
with a semicolon denoting a covariant derivative in the field space. Therefore, we
obtain
\begin{equation}\label{multirrunning}
\frac{d\log r}{d\log k} - \frac{r}{4} = -2\eta_\mathrm{multi} +
2\frac{N_{,i}N_{,j}}{h^{kl}N_{,k}N_{,l}}
\frac{{R^i_{\,\,ab}}^j}{3m_\mathrm{Pl}^2}\frac{\dot\phi^a\dot\phi^b}{H^2} \, ,
\end{equation}
where on the left hand side are only observable quantities. We can see that from
Eqs.~(\ref{gwindex}) and (\ref{multiPindex}) the common term $-2\epsilon$ disappears
and we are left with the ones which depend purely on the geometry of the inflaton
field space and the inflationary dynamics. It is very important to note that in
contrast to the single field case, we cannot obtain these terms with only
$\mathcal{P}_\mathrm{S}$, $\mathcal{P}_\mathrm{T}$ and $n_\mathrm{S}$. Therefore
measuring the running of the tensor-to-scalar ratio can be another way of
distinguishing between single and multi-field cases observationally, and further
provide a potential probe of the underlying theory of inflation.

An immediate non-trivial contribution appears when the field space is curved. For
example, let us consider the K{\"a}hler potential
\begin{equation}
K = -3m_\mathrm{Pl}^2 \log \left( T + T^* \right) \, ,
\end{equation}
which is typical for string moduli. Writing the complex modulus field $T$ as a sum
of two real fields $X$ and $Y$, i.e. $T = X + i Y$, from the kinetic term
\begin{equation}
\frac{\partial^2K}{\partial\Phi^I\partial\Phi^{*J}}
\partial^\mu\Phi^I\partial_\mu\Phi^{*J} = \frac{1}{2}h_{ij}
\partial^\mu\varphi^i\partial_\mu\varphi^j \, ,
\end{equation}
we find the metric $h_{ij}$ for the real scalar fields $\varphi^i$ as
\begin{equation}
h_{ij} = \mathrm{diag}\left( \frac{3m_\mathrm{Pl}^2}{2X^2},
\frac{3m_\mathrm{Pl}^2}{2X^2} \right) \, .
\end{equation}
Then we have
\begin{equation}
R^X_{\,\,\,YXY} = R^Y_{\,\,\,XYX} = \frac{1}{X^2} \, ,
\end{equation}
so the field space is indeed curved, and leads to additional contributions on the
right hand side of Eq.~(\ref{multirrunning}). However, it is expected that this
contribution due to curved field space is small: generally, in the slow-roll regime,
the variation of the field $\phi_i$ per each $e$-fold is
\begin{equation}
\frac{\Delta\phi_i}{m_\mathrm{Pl}} = \sqrt{2\epsilon_i} \, ,
\end{equation}
with $\epsilon_i \equiv (\dot\phi_i/H)^2/(2m_\mathrm{Pl}^2)$ so that $\epsilon =
\sum_i \epsilon_i$. This means while the scales corresponding to the low multipoles
$l \lesssim 100$ are leaving the horizon, the probed field space is very tiny.
Therefore whatever geometry the field space has actually, it can be approximated as
a flat space.

\begin{figure}[h]
\begin{center}
\epsfig{file = 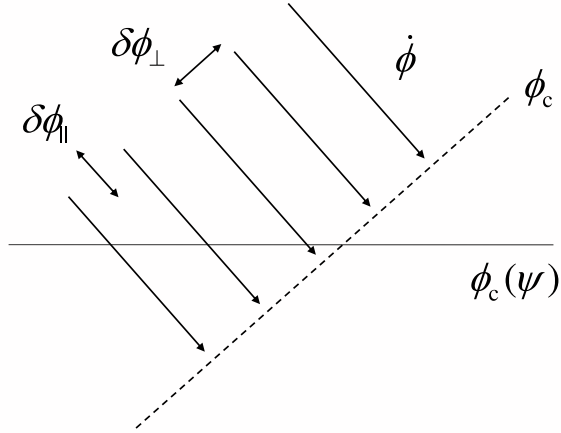, angle = 90, width = 8cm}
\end{center}
\caption{A simple trajectory where the critical value of the inflaton field $\phi$
is {\em not} a constant. Here $\delta{N}$ depends not only on $\delta\phi_{||}$, the
fluctuations along $\dot\phi$, making $\eta_\mathrm{multi}$ non-trivial.}
\label{hybridtraj}
\end{figure}

However, even if the field space is flat, i.e. $h_{ij} = \delta_{ij}$, in general we
still have non-trivial result. This is because $V^{;ij}$ is contracted with
$N_{,i}N_{,j}$, i.e. we have a `projection' operator. From $dN = H dt$ it directly
follows that
\begin{equation}
N_{,i} \dot\phi^i = \nabla{N}\cdot\dot\Phi = H \, ,
\end{equation}
which {\em never} necessarily means that the gradient of $N$ in the field space is
aligned in a specific manner with respect to the corresponding component of the
velocity of the field. For example, consider a simple hybrid like situation depicted
in Fig.~\ref{hybridtraj}. In usual hybrid inflation~\cite{hybrid}, $\phi_\mathrm{c}$
is a constant and is orthogonal to the direction of the evolution of $\phi$, i.e.
$\dot\phi$. Thus $\delta\phi_\bot$, the fluctuations of $\phi$ orthogonal to
$\dot\phi$, cannot alter the number of $e$-folds $N$ and hence the curvature
perturbation, which is equivalent to $\delta{N}$, is independent of
$\delta\phi_\bot$. Only the fluctuations along $\dot\phi$, which we write
$\delta\phi_{||}$, can change the end of inflation by moving $\phi$ backward or
forward by that amount, and accordingly give rise to $\delta{N}$ and the curvature
perturbation. Now, instead of a constant $\phi_\mathrm{c}$, let us consider the case
where $\phi_\mathrm{c}$ is not orthogonal to $\dot\phi$ due to for example another
field $\psi$, i.e. $\phi_\mathrm{c} = \phi_\mathrm{c}(\psi)$. In this case,
$\delta{N}$ is not only dependent on $\delta\phi_{||}$, but also on
$\delta\phi_\bot$. Therefore $\nabla{N}$ does {\em not} have a simple relation with
$\dot\phi$ such as
\begin{equation}
\frac{\partial{N}}{\partial\phi} = \frac{H}{\dot\phi} \, ,
\end{equation}
which holds for single field case. Contracted with $N_{,i}$, even in flat field
space generally we have different result for $m_\mathrm{Pl}^2
(N_{,i}N_{,j}/N_{,k}N^{,k}) V^{,ij}/V$ from single field case where this corresponds
to simply $\eta$. This suggests the potential role of the running of $r$ as an
observational probe of the non-trivial inflationary dynamics. The simplest multiple
chaotic inflation model,
\begin{equation}
V = \sum_i V_i = \sum_i \lambda_i \frac{\phi_i^n}{m_\mathrm{Pl}^{n-4}} \, ,
\end{equation}
is in fact a very special case~\cite{multichaotic} where we can reproduce the
predictions of the corresponding single field case
\begin{equation}
V = \lambda \frac{\phi^n}{m_\mathrm{Pl}^{n-4}} \, .
\end{equation}

Finally, let us discuss the practical feasibility of the detection of
$d\log{r}/d\log{k}$. This amounts to, as can be read from Eqs.~(\ref{rrunning}),
(\ref{singlerrunning}) and (\ref{multirrunning}), the individual detection of
$n_\mathrm{S}$, $n_\mathrm{T}$ and $r$. Apart from $n_\mathrm{S}$ which will be even
further constrained, the $B$-mode CMB polarization anisotropies can reveal the
primordial gravitational waves and thus $r$ and $n_\mathrm{T}$. By the CMB
polarization observations alone, e.g. even CMBPol, unfortunately, it is unlikely
that we can determine both of them, or at least $n_\mathrm{T}$, at a satisfactory
level: for a realistic experiment with no foreground and no lensing subtraction
which can probe $r \gtrsim 0.01$, the $1\sigma$ error of $n_\mathrm{T}$ is as large
as $\mbox{few} \times 0.01 - 0.1$~\cite{cmbpolobs}, but $n_\mathrm{T}$ given by
Eq.~(\ref{gwindex}) is at most as large as $\mathcal{O}(0.01)$. This situation,
however, can be greatly improved by direct detection programmes with laser
interferometers such as Big Bang Observer (BBO) and Deci-Hertz Interferometer
Gravitational wave Observer (DECIGO). Their sensitivity peaks are at around 1 Hz
with $\Omega_\mathrm{GW}h^2 \sim 10^{-18} \, \mbox{(BBO)} - 10^{-20} \,
\mbox{(DECIGO)}$, which corresponds to $r \sim 10^{-4} \, \mbox{(BBO)} - 10^{-6} \,
\mbox{(DECIGO)}$. Then, the $1\sigma$ error of $n_\mathrm{T}$ can be improved as
much as $\sigma_{n_\mathrm{T}} \sim 10^{-8}/r$ in this
regime~\cite{gwinterferometer}. Thus from the laser interferometer experiments, it
is not completely impossible to detect $d\log{r}/d\log{k}$ provided that once $r$ is
detected. With the sharpening accuracy of the observations of the scalar
perturbations on the CMB scales, it could be sufficient enough with laser
interferometer alone: if $r$ is detected on the CMB scales, we will be able to fix
everything unambiguously.

To conclude, we have found a very simple consistency relation from the running of
the tensor-to-scalar ratio first order in the slow-roll parameters. In single field
models this would merely reproduce what we can find by using other observables,
meanwhile for multi-field cases we can extract very important information on the
inflaton field space and the inflationary dynamics observationally. We can hope to
detect this running by the future laser interferometer experiments.

I am indebted to David Lyth, Misao Sasaki, Ewan Stewart and Takahiro Tanaka for
important comments and suggestions. I am also deeply grateful to Donghui Jeong and
Yong-Seon Song for crucial discussions about the practical feasibility of various
observations. I acknowledge support while this work was in progress from the Yukawa
Institute for Theoretical Physics at Kyoto University during ``Scientific Program on
Gravity and Cosmology'' (YITP-T-07-01) and ``KIAS-YITP Joint Workshop: String
Phenomenology and Cosmology'' (YITP-T-07-10). I am also partly supported by the
Korea Research Foundation Grant KRF-2007-357-C00014 funded by the Korean Government.


\begin{thebibliography}{99}


\bibitem{inflation}
A.~H. Guth, \textit{Phys. Rev.} \textbf{D 23}, 347 (1981)~; A.~D. Linde,
\textit{Phys. Lett.} \textbf{B 108}, 389 (1982)~; A. Albrecht and P.~J. Steinhardt,
\textit{Phys. Rev. Lett.} \textbf{48}, 1220 (1982)


\bibitem{books}
See, e.g. A.~D. Linde, \textit{Particle physics and inflationary cosmology}, Harwood
Academic Press (1990)~; A.~R. Liddle and D.~H. Lyth, \textit{Cosmological inflation
and large scale structure}, Cambridge University Press (2000)~; V.~F. Mukhanov,
\textit{Physical foundations of cosmology}, Cambridge University Press (2005)


\bibitem{cobe}
G.~F. Smoot et al., \textit{Astrophys. J.} \textbf{396}, L1 (1992)


\bibitem{observations}
M. Tegmark et al., \textit{Phys. Rev.} \textbf{D 74}, 123507 (2006)
\texttt{astro-ph/0608632}~; J.~K. Adelman-McCarthy et al., \textit{Astrophys. J.
Suppl.} \textbf{172}, 634 (2007) \texttt{0707.3380 [astro-ph]}~; E. Komatsu et al.,
\texttt{0803.0547 [astro-ph]}


\bibitem{gw}
A.~A. Starobinsky, \textit{JETP Lett.} \textbf{30}, 682 (1979)


\bibitem{infE}
L. Knox and Y.-S. Song, \textit{Phys. Rev. Lett.} \textbf{89}, 011303 (2002)
\texttt{astro-ph/0202286}


\bibitem{kklt}
S. Kachru, R. Kallosh, A. Linde and S.~P. Trivedi, \textit{Phys. Rev.} \textbf{D
68}, 046005 (2003) \texttt{hep-th/0301240}


\bibitem{stringgw}
D. Baumann and L. McAllister, \textit{Phys. Rev.} \textbf{D 75}, 123508 (2007)
\texttt{hep-th/0610285}


\bibitem{lythbound}
D.~H. Lyth, \textit{Phys. Rev. Lett.} \textbf{78}, 1861 (1997)
\texttt{hep-ph/9606387}


\bibitem{largeGW}
See, however, for example A. Krause, \textit{J. Cosmol. Astropart. Phys.}
\textbf{07}, 001 (2008) \texttt{arXiv:0708.4414 [hep-th]}


\bibitem{gwcalc}
J.-O. Gong, \textit{Class. Quant. Grav.} \textbf{21}, 5555 (2004)
\texttt{gr-qc/0408039}


\bibitem{curvaton}
D.~H. Lyth and D. Wands, \textit{Phys. Lett.} \textbf{B 524}, 5 (2002)
\texttt{hep-ph/0110002}~; T. Moroi and T. Takahashi, \textit{Phys. Lett.} \textbf{B
522}, 215 (2001) \texttt{hep-ph/0110096}~; \textit{Erratum-ibid.} \textbf{B 539},
303 (2002)


\bibitem{postinfr}
J.-O. Gong, \textit{Phys. Lett.} \textbf{B 657}, 165 (2007) \texttt{arXiv:0706.3599
[astro-ph]}


\bibitem{singlecorr}
E.~D. Stewart and J.-O. Gong, \textit{Phys. Lett.} \textbf{B 510}, 1 (2001)
\texttt{astro-ph/0101225}


\bibitem{generalSR}
S. Dodelson and E. Stewart, \textit{Phys. Rev.} \textbf{D 65}, 101301(R) (2002)
\texttt{astro-ph/0109354}~; E.~D. Stewart, \textit{Phys. Rev.} \textbf{D 65}, 103508
(2002) \texttt{astro-ph/0110322}~; J. Choe, J.-O. Gong and E.~D. Stewart, \textit{J.
Cosmol. Astropart. Phys.} \textbf{07}, 012 (2004) \texttt{hep-ph/0405155}


\bibitem{hierarchy}
M. Cort\^{e}s and A.~R. Liddle, \textit{Phys. Rev.} \textbf{D 73}, 083523 (2006)
\texttt{astro-ph/0603016}


\bibitem{zoology}
S. Dodelson, W.~H. Kinney and E.~W. Kolb, \textit{Phys. Rev.} \textbf{D 56}, 3207
(1997) \texttt{astro-ph/9702166}


\bibitem{deltaN}
A.~A. Starobinsky, \textit{JETP Lett.} \textbf{42}, 152 (1985)~; M. Sasaki and E.~D.
Stewart, \textit{Prog. Theor. Phys.} \textbf{95}, 71 (1996)
\texttt{astro-ph/9507001}~; M. Sasaki and T. Tanaka, \textit{Prog. Theor. Phys.}
\textbf{99}, 763 (1998) \texttt{gr-qc/9801017}~; J.-O. Gong and E.~D. Stewart,
\textit{Phys. Lett.} \textbf{B 538}, 213 (2002) \texttt{astro-ph/0202098}~; D.~H.
Lyth, K.~A. Malik and M. Sasaki, \textit{J. Cosmol. Astropart. Phys.} \textbf{05},
004 (2005) \texttt{astro-ph/0411220}


\bibitem{hybrid}
A. Linde, \textit{Phys. Rev.} \textbf{D 49}, 748 (1994) \texttt{astro-ph/9307002}


\bibitem{multichaotic}
L. Alabidi and D.~H. Lyth, \textit{J. Cosmol. Astropart. Phys.} \textbf{05}, 016
(2006) \texttt{astro-ph/0510441}~; J.-O. Gong, \textit{Phys. Rev.} \textbf{D 75},
043502 (2007) \texttt{hep-th/0611293}


\bibitem{cmbpolobs}
Y.-S. Song and L. Knox, \textit{Phys. Rev.} \textbf{D 68}, 043518 (2003)
\texttt{astro-ph/0305411}~; L. Verde, H.~V. Peiris and R. Jimanez, \textit{J.
Cosmol. Astropart. Phys.} \textbf{01}, 019 (2006) \texttt{astro-ph/0506036}


\bibitem{gwinterferometer}
T.~L. Smith, M. Kamionkowski and A. Cooray, \textit{Phys. Rev.} \textbf{D 73},
023504 (2006) \texttt{astro-ph/0506422}~; H. Kudoh, A. Taruya, T. Hiramatsu and Y.
Himemoto, \textit{Phys. Rev.} \textbf{D 73}, 064006 (2006) \texttt{gr-qc/0511145}~;
T.~L. Smith, H.~V. Peiris and A. Cooray, \textit{Phys. Rev.} \textbf{D 73}, 123503
(2006) \texttt{astro-ph/0602137}


\end{thebibliography}
\end{document}